# Probing Dynamics at Interfaces: Molecular Motions in Lipid Bilayers studied by Neutron Backscattering

Maikel C. Rheinstädter and Tilo Seydel (ILL)
Tim Salditt (Institut für Röntgenphysik, Georg-August Universität Göttingen, Germany)

*Lipid membranes in a physiological context cannot be understood without taking into account their mobile environment. Here, we report on a high energy-resolution neutron backscattering study to investigate slow motions on nanosecond time scales in highly oriented solid supported phospholipid bilayers of the model system DMPC -d54 (deuterated 1,2-dimyristoyl-sn-glycero-3-phoshatidylcholine). This technique allows discriminating the Q-dependent onset of mobility and provides a benchmark test regarding the feasibility of dynamical neutron scattering investigations on these sample systems. Apart from freezing of the lipid acyl-chains, we could observe a second freezing temperature that we attribute to the hydration water in between the membrane stacks. The freezing is lowered several degrees as compared to (heavy) bulk water.*

Lipid membranes as model systems for more complex biological membranes (1) cannot be understood without taking into account the structure and dynamics of their aqueous environment. The structure and dynamical properties of the bound water layers next to the bilayer as well as the 'free' (bulk) water further away from the water/lipid interface are of great importance in understanding the thermal, elastic and transport properties of the membrane. While most spectroscopic techniques such as nuclear magnetic resonance or dielectric spectroscopy, are limited to the centre of the Brillouin zone at $Q=0$ and probe the macroscopic response of the system, neutrons and within some restrictions also X-rays give unique access to microscopic dynamics at length scales of e.g. intermolecular distances. Here, we report on a high energy-resolution (sub-$\mu$eV) neutron backscattering study to investigate slow molecular motions on nanosecond time scales in highly oriented solid supported phospholipid bilayers of the model system DMPC -d54 (deuterated 1,2-dimyristoyl-sn-glycero-3-phoshatidylcholine), hydrated with heavy water. The scattering volume restriction resulting from the low scattering volume of quasi-two dimensional planar membranes and the small inelastic signal was overcome by stacking several thousand highly aligned membrane bilayers.

The experiment was carried out at the cold neutron backscattering spectrometer IN10 in its standard setup with Si(111) monochromator and analyzer crystals corresponding to an incident and analyzed neutron energy of 2.08 meV ($\lambda$=6.27 Å). Two types of measurements have been performed: With fixed energy-window scans centred at zero energy transfer (FEW-scans), the scattered intensity arising from the sample, which is elastic within the instrumental resolution, was recorded as a function of the sample temperature. From FEW-scans, information on the onset and type of molecular mobility in the sample can be inferred. Thus, glass or melting transitions can be clearly identified and assigned to corresponding length scales by analyzing the corresponding $Q$-dependence. The second type of measurement was performed by Doppler-shifting the incident neutron energy through an adequate movement of the monochromator crystal.

The IN10 analyzers cover an angular range of approximately 20 ° each, resulting in a rather poor $Q$-resolution, but enhanced sensitivity for even very small inelastic signals. We used six discrete detector tubes of IN10. The broad lipid acyl-chain correlation peak that occurs at $Q_{\cong}1.4$ Å$^{-1}$ was (mainly) detected in one detector tube ('lipid detector'), as depicted in Fig. 1. A Q-range of 0.3 Å$^{-1}$<$Q$<1.9 Å$^{-1}$ was simultaneously detected in this set-up to investigate and discriminate molecular dynamics on the different length scales.

We performed FEW scans in a temperature range of 100-315 K to map out the transition of the lipids from immobile to mobile as a function of temperature for (a) the scattering vector $Q$ placed in the plane of the membranes and (b) perpendicular to the bilayers. While the in-plane component ($Q_r$) in the 'lipid-detector' shows a pronounced freezing transition, there is no distinct T-dependence in the perpendicular direction ($Q_z$). We interpret this in terms of correlated motions, which take place mainly in the plane of the lipid bilayers (in the time and length scales observed). Figure 2 shows the in-plane component of the elastic scattering with the measurement in $Q_z$ subtracted as background. We attribute the pronounced freezing step ('immobile' within the resolution window) at 294 K ($Q$ centred at 1.42 Å$^{-1}$) to the main transition of the lipid acyl-chains from the rigid gel phase at low-T into the fluid phase at higher temperatures. When analysing all detectors we find a second transition at about 271 K, mainly in the detector centred at $Q$=1.85 Å$^{-1}$, which tentatively might be attributed to the hydration water of the membrane stacks, i.e. the water layer in between the stacked membranes. Even though the detector is not perfectly centred to the maximum of the static structure factor of water at $Q$=2 Å$^{-1}$ (which is not accessible on IN10), it is positioned to detect a reasonable part of the broad heavy water correlation peak. Freezing of the hydration water, hence, is lowered by about six degrees as compared to (heavy) bulk water at about 277 K. Fig.3 displays corresponding energy transfer scans. The data have been taken at three different temperatures, for T=250, 290 and 300 K with a typical counting time of about 9 hours per temperature. An elastic peak in the inelastic spectra points to static order at the corresponding length scales, where a fluid system has no order at infinitely long time scales. Even within the very limited statistics, the different dynamics is clearly visible: While the lipid acyl-chains melt between 290 and 300 K, melting at the water position already occurs between 250 and 290 K.

Our experiment gives a first high energy-resolution wave vector-resolved insight into collective lipid membrane dynamics. The dynamical properties of hydration water may be different from those of bulk water because hydrogen bonding to the lipid head groups at the lipid-water interface of the membrane might slow down water rotation and translation (3). A scenario with gradual freezing of the water molecules, depending on the distance to the water-lipid interface, is under discussion. Only recently, Tarek and Tobias got access to the single-particle and collective dynamics of hydration water of a protein (4) by Molecular Dynamics (MD) simulations pointing out the importance of water dynamics for the understanding of the dynamical transition of the protein. The membrane-water dynamics must play a crucial role for the understanding of fluctuations in stacked membrane systems as the hydration water e.g. mediates the interactions between two bilayers (5).

(1) R. Lipowsky, R. and E. Sackmann, Structure and Dynamics of Membranes, Handbook of Biological Physics 1 (1995) Elsevier, North-Holland, Amsterdam.
(2) M.C. Rheinstädter, C. Ollinger, G. Fragneto, F. Demmel and T. Salditt, Phys. Rev. Lett. 93, 108107 (2004).
(3) M. Settles and W. Doster, Faraday Discuss. 103, 269 (1996).
(4) M. Tarek and D. Tobias, Phys. Rev. Lett. 88, 138101 (2002); M. Tarek and D. Tobias, Phys. Rev. Lett. 89, 275501 (2002).
(5) J. N. Israelachvili and H. Wennerstroem, Langmuir 6, 873 (1990); L. Perera, U. Essmann, and M. Berkowitz, Langmuir 12, 2625 (1996); J. Katsaras and K. R. Jeffrey, Europhys. Lett. 38, 43 (1997)

**Figure 1:.**Schematic of the scattering geometry. The inter-acyl-chain correlation peak in the plane of the membranes is located at 1.4 Å$^{-1}$ (the heavy water correlation peak occurs at 2 Å$^{-1}$). Spatially arranged analyzers allow to separately but simultaneously probe the molecular dynamics at different length scales.

**Figure 2:** In-plane component of the elastic scattering signal. The mobile-immobile transition (within the resolution window) is clearly different for the lipid acyl-chains (T=294 K) and the position of the water correlation peak (T=271 K). Solid lines are guides to the eye. (Counting is normalized to monitor)

**Figure 3:.**Energy scans at temperatures T=250 K, 290 K and 300 K for the *Q*-values 1.42 Å$^{-1}$ (lipid acyl-chain correlation peak) and *Q*=1.85 Å$^{-1}$. At 290 K, the water signal is already 'mobile' within the experimental energy resolution whereas the lipid acyl-chains are still frozen. (Counting is normalized to monitor)

Figure 1

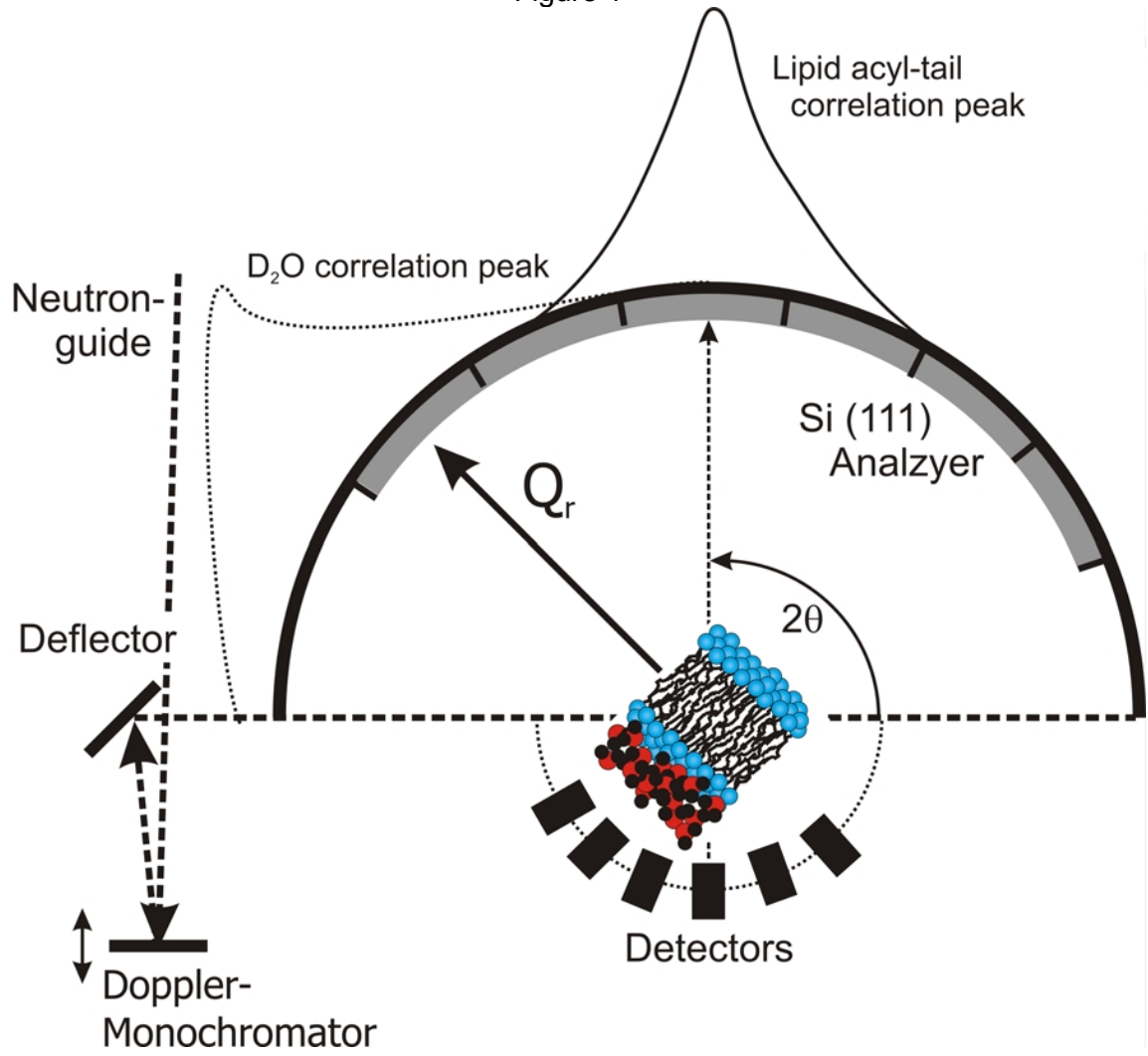

Lipid acyl-tail correlation peak

$D_2O$ correlation peak

Neutron-guide

Si (111) Analzyer

$Q_r$

$2\theta$

Deflector

Doppler-Monochromator

Detectors

Figure 2

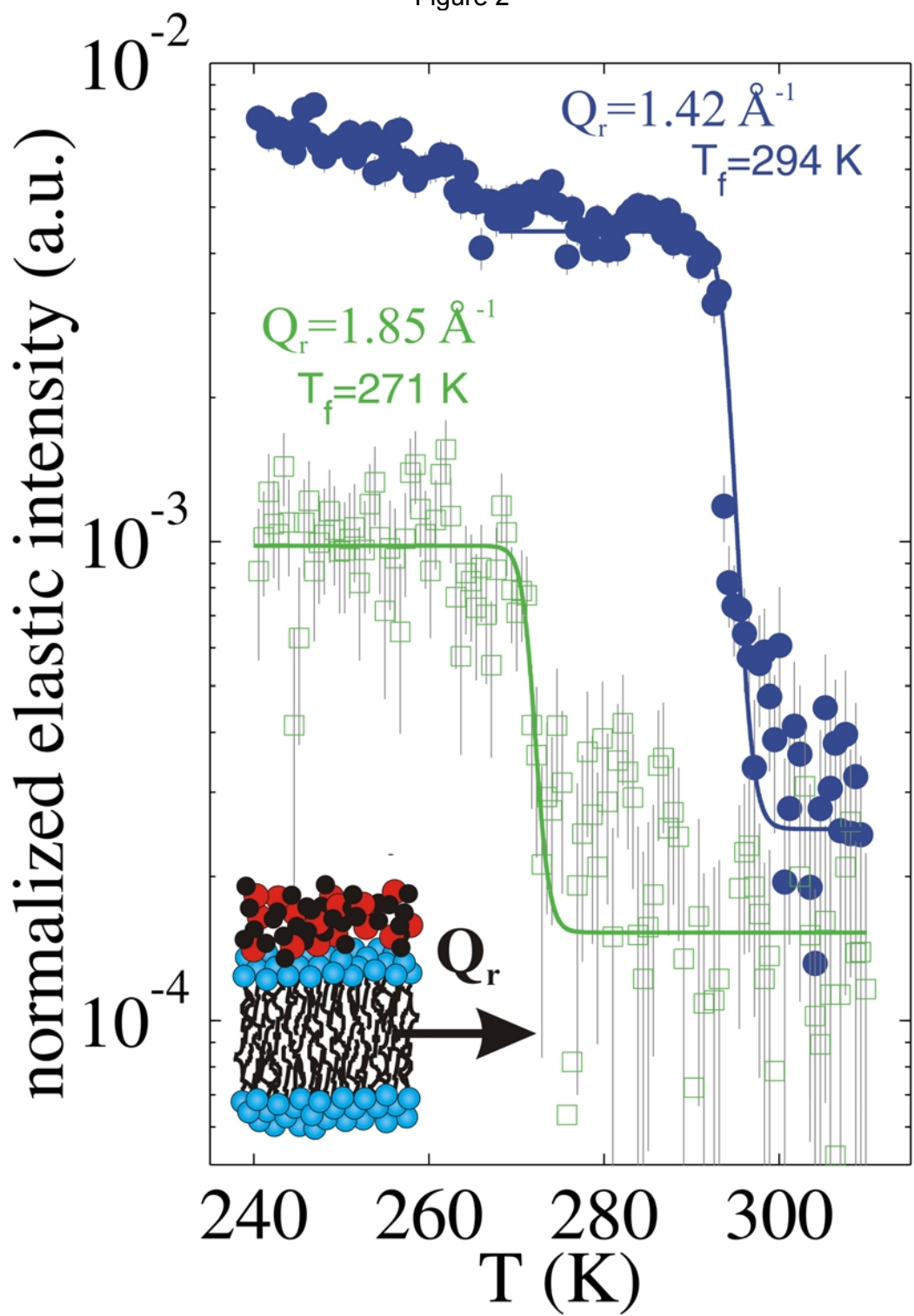

Figure 3

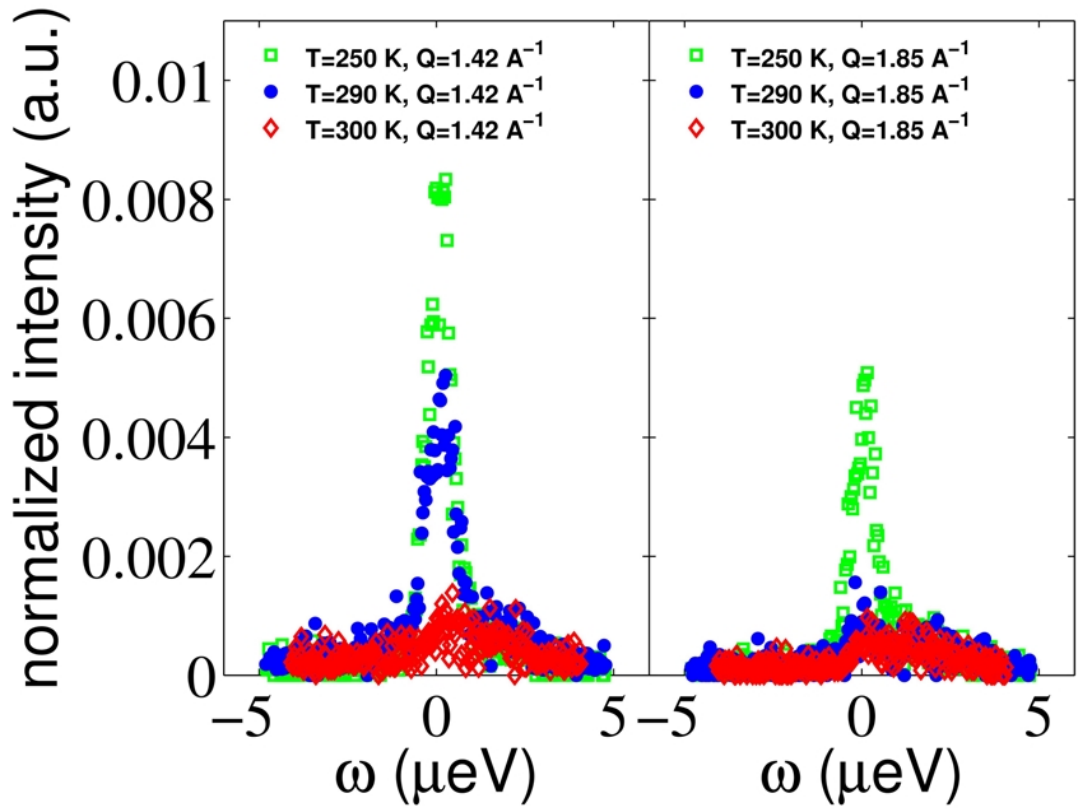